\documentclass[10pt,letterpaper]{article}
\usepackage{opex3}
\usepackage{subfigure}
\usepackage{amsmath}
\usepackage{graphicx}
\usepackage{epstopdf}
\usepackage{cite}

\begin{document}

\title{Phase noise in collective binary phase shift keying with Hadamard words}

\author{Marcin Jarzyna,$^1$ Victoria Lipi\'{n}ska,$^1$ Aleksandra Klimek,$^1$ Konrad~Banaszek,$^{1,*}$
and Matteo G. A. Paris$^2$}

\address{$^1$Faculty of Physics, University of Warsaw, Pasteura 5, PL-02-093 Warsaw, Poland\\
$^2$Dipartimento di Fisica, Universit\`{a} degli Studi di Milano, I-20133 Milano, Italy}

\email{\textsuperscript{*}Konrad.Banaszek@fuw.edu.pl} 



\begin{abstract}
We analyze the effect of phase fluctuations in an optical communication scheme based on collective detection of sequences of binary coherent state symbols using linear optics and photon counting. When the phase noise is absent, the scheme offers qualitatively improved nonlinear scaling of the spectral efficiency with the mean photon number in the low-power regime compared to individual detection. We show that this feature, providing a demonstration of superaddivitity of accessible information in classical communication over quantum channels, is preserved if random phases imprinted on transmitted symbols fluctuate around a reference fixed over the sequence length.
\end{abstract}

\ocis{(270.5565) Quantum communications; (030.5290) Photon statistics; (060.4510) Optical communications.} 



\section{Introduction}
Binary phase shift keying (BPSK) is a well established modulation format in optical comunication \cite{Kahn,Kitayama}. It employs a pair of coherent states with the same mean photon number $\bar{n}$ and opposite phases $+ \equiv e^{i0}$ and $- \equiv e^{i\pi}$. In the low-power regime, when $\bar{n} \ll 1$, its readout becomes non-trivial because of quantum fluctuations, which prevent perfect discrimination between the two symbols constituting the BPSK alphabet. For individual detection of BPSK symbols, the minimum discrimination error is given by the celebrated Helstrom bound and can be  attained using the Dolinar receiver \cite{DolinarProposal,DolinarExp} based on photon counting and fast feed-forward operations. The resulting spectral efficiency, given by Shannon mutual information expressed in bits, scales in the low-power regime linearly with the mean photon number as
$I_{\text{Hel}} \approx ( 2 / \ln 2) \bar{n}$. Quantum mechanics offers a possibility to change qualitatively the above scaling through collective detection of multiple symbols. The ultimate limit of such a general strategy is given by the Holevo quantity $\chi$ \cite{Holevo} which in the case of BPSK expands for $\bar{n} \ll 1$ as $\chi \approx  \bar{n} \log_2 (1/\bar{n})$, exhibiting enhanced nonlinear scaling with the mean photon number. The ability to improve the attainable transmission rate by implementing a joint measurement on sequences of symbols is referred to in quantum communication theory as the {\em superadditivity} of accessible information \cite{SasakiPRA}.

Recently, a linear-optics scheme to attain superadditivity for the BPSK format has been proposed by Guha \cite{Guha}. The basic idea is to prepare sequences of BPSK symbols defined by columns of a Hadamard matrix. Because of orthogonality properties of Hadamard matrices, applying an appropriate linear-optics transformation to such Hadamard words maps them onto the pulse position modulation (PPM) format that is suitable for direct detection. This strategy can approach in the leading order the Holevo quantity when optimized over the sequence length \cite{Kochman,Jarzyna}.
The purpose of this paper is to study the impact of excess phase noise on collective BPSK with Hadamard words. As the main result, we demonstrate that the nonlinear scaling of spectral efficiency in the low-power regime is retained when individual BPSK symbols experience small phase fluctuations occurring around a reference locked over the sequence length. Remarkably, the enhanced scaling persists despite the optimal length of Hadamard words growing indefinitely large as the mean photon number tends to zero.

This paper is organized as follows. In Sec.~\ref{Sec:Collective} we review collective BPSK with Hadamard words and introduce the phase noise model. Optimization of the spectral efficiency is considered in Sec.~\ref{Sec:Optimization}. Finally, Sec.~\ref{Sec:Conclusions} concludes the paper.


\section{Collective BPSK with Hadamard words}
\label{Sec:Collective}

\begin{figure*}
   \centering
   \includegraphics[width=\columnwidth]{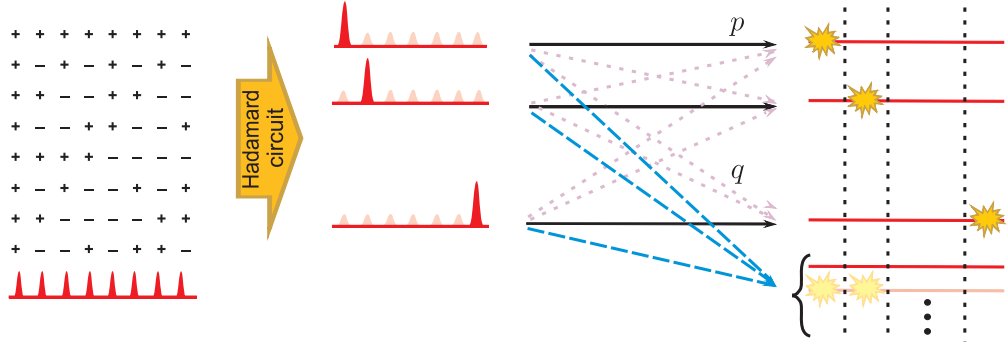}
    \caption{Sequences of BPSK symbols $\pm$ shown on the left are prepared as rows (or equivalently columns) of a symmetric Hadamard matrix. A linear circuit, described by a rescaled Hadamard matrix, transforms them into the PPM format visualized as tall solid red pulses localized in single output bins. When sequences are subject to phase noise, some of the input sequence energy becomes evenly distributed across all other bins, depicted with light red pulses. Assuming no dark counts, the sequence is either identified correctly with probability $p$ by the position of the detector click (solid black arrows) or erased when no photocounts are generated (dashed blue arrows). In the presence of phase noise, the fraction of the sequence energy spread over all remaining PPM bins may produce a click in a given wrong bin with a probability $q$ (dotted purple arrows). Events when clicks occur in two or more bins are treated as erasures.}
    \label{Fig:Hadamardwords}
\end{figure*}

Hadamard matrices are square symmetric matrices with entries $h_{jl} = h_{lj} = \pm 1$ and mutually orthogonal columns, or equivalently rows. They exist for dimensions $L$ equal to powers of $2$ and selected other integers. In the proposed collective BPSK scheme \cite{Guha} shown in Fig.~\ref{Fig:Hadamardwords}, sequences of input symbols, called in the following {\it Hadamard words}, are chosen as columns of a Hadamard matrix, i.e.\ the $l$th word is given by $(h_{1l}, h_{2l}, \ldots, h_{Ll})$, where $l=1,2,\ldots,L$. These symbols define the complex amplitudes of individual pulses, i.e.\ the amplitude of the $j$th pulse in the $l$th word is $h_{jl}\sqrt{\bar{n}}$, where $\bar{n}$ is the mean photon number per time bin.

At the detection stage the sequence, after synchronizing the time bins, is fed into a linear optics circuit that implements a unitary transformation of input complex amplitudes described by a rescaled Hadamard matrix $(h_{kj}/\sqrt{L})$. The output amplitude at the $k$th port is thus given by $\sqrt{\bar{n}/L} \sum_{j} h_{kj} h_{jl} = \sqrt{L\bar{n}}\delta_{kl}$, where in the second expression we have made use of the orthogonality properties of Hadamard matrices and $\delta_{kl}$ denotes Kronecker delta. It is seen that amplitudes are zeroed at all output ports of the circuit except the $l$th one, which carries the entire sequence energy characterized by the mean photon number $L\bar{n}$. The non-empty port, and consequently the input Hadamard sequence, can be identified using photodetectors monitoring individual output ports. Neglecting dark counts and assuming that binary on/off detectors are used, identification is unambiguous when one of the detectors generates a photocount. In the absence of any detection event, no information about the input sequence is recovered.

From the communication theory viewpoint the above scheme is described by an erasure channel \cite{CoverThomas} with $L$ equiprobable values of the input variable that is either transmitted faithfully through the channel or replaced by an erasure flag. In our case, the erasure probability is given by $\exp(-L\bar{n})$, which corresponds to registering no photocount at any output port. For such a scenario, Shannon mutual information per time bin reads $L^{-1} [1- \exp(-L\bar{n})] \log_2 L$. When $\bar{n} \ll 1 $, the optimal sequence length can be found approximately by expanding $1-\exp(-L\bar{n}) \approx L\bar{n} - (L\bar{n})^2/2$ \cite{Kochman,Jarzyna}. This calculation yields spectral efficiency with the same leading-order dependence on $\bar{n}$ as the Holevo quantity $\chi\approx  \bar{n} \log_2 (1/\bar{n})$.

Suppose now that the transmission of each BPSK symbol is affected by uncorrelated Gaussian phase noise\cite{Olivares} characterized by a standard deviation $\sigma$. In this model, before the readout stage the phases of individual BPSK symbols are $(e^{i\phi_1} h_{1l}, e^{i\phi_2} h_{2l}, \ldots, e^{i\phi_L} h_{Ll})$, where $L$ random phase variables $\phi_1, \phi_2, \ldots, \phi_L$ are described by the same Gaussian distribution
\begin{equation}
p(\phi) = \exp[-\phi^2/(2\sigma^2)]/\sqrt{2\pi\sigma^2}.
\end{equation}
Because of the phase noise, at the output of the Hadamard circuit the energy of the sequence will no longer be concentrated at a single output port. Consequently, the communication scheme cannot be represented as a standard erasure channel, but a possibility of incorrect identification of the input sequence also needs to be taken into account.

Let us analyze the above more general scenario in quantitative terms. For the $l$th Hadamard word and an individual realization of all random phase variables, the $k$th output port of the linear optics circuit carries a coherent state with an average photon number
\begin{equation}
\mu_{k}^{(l)} = \frac{\bar{n}}{L} \left| \sum_{j=1}^{L} e^{i\phi_j} h_{kj} h_{jl} \right|^2.
\label{Eq:mukl}
\end{equation}
We will be interested only in events when exactly one detector clicks. Events when two or more detectors produce photocounts will be treated as erasures. The exclusive probability for the detector monitoring the $k$th port to generate at least one photocount, while all other do not click at all, reads
\begin{equation}
p_{k}^{(l)}  = \left\langle (1 - e^{-\mu^{(l)}_{k}}) \prod_{j \neq k } e^{-\mu^{(l)}_{j}} \right\rangle
=
\left\langle (e^{\mu^{(l)}_{k}} - 1  ) \prod_{j } e^{-\mu^{(l)}_{j}} \right\rangle
=
e^{-L\bar{n}} \bigl( \langle e^{\mu^{(l)}_{k}} \rangle - 1 \bigr),
\label{Eq:pjlgeneral}
\end{equation}
where the angular brackets denote the statistical average over the phase noise. The last  simplified expression for $p_{k}^{(l)}$ in Eq.~(\ref{Eq:pjlgeneral}) is obtained from energy conservation in the course of the linear circuit transformation, which implies that $\sum_{j=1}^{L} \mu^{(l)}_{j} = L\bar{n}$. It is easy to see that the probability $p_{k}^{(l)}$ assumes one of only two values depending on whether $k=l$ or $k \neq l$. This is because in Eq.~(\ref{Eq:mukl}) all the products $h_{kj} h_{jl}$ are equal to one when $k=l$ and they are split evenly between $1$ and $-1$ when $k \neq l$, the latter property following from the orthogonality of Hadamard matrix columns. Consequently, it is sufficient to introduce two parameters: the probability of correctly identifying the input Hadamard word $p=p_{l}^{(l)}$ and the probability of obtaining a wrong answer, $q=p_{k}^{(l)}$ for $k\neq l$. The resulting communication scheme can be viewed as a generalized erasure channel with uniform noise also depicted in Fig.~\ref{Fig:Hadamardwords}, which in addition to deleting the input variable can also scramble its value.

The formulas for the parameters $p$ and $q$ following from Eq.~(\ref{Eq:pjlgeneral}) involve arbitrarily high moments of the distributions of random phase variables $\phi_j$ and are difficult to handle analytically. One can however derive simple bounds on $p$ and $q$ that will prove useful when optimizing the sequence length. The starting point is to
average over the phase noise the mean photon number at individual output ports. This yields
\begin{equation}
\langle \mu^{(l)}_{k} \rangle = \bar{n} [ 1+ (L\delta_{kl} -1) e^{-\sigma^2}].
\label{Eq:avgmu}
\end{equation}
In the case of $p$, the convexity of the exponential function implies that
$\langle e^{\mu^{(l)}_{l}} \rangle \ge e^{\langle \mu^{(l)}_{l}\rangle} $. Using Eq.~(\ref{Eq:avgmu}) for $k=l$ yields
\begin{equation}
p  \ge  \exp[-(L-1)\bar{n}(1-e^{-\sigma^2})] - \exp(-L\bar{n}).
\label{Eq:papprox}
\end{equation}
On the other hand, the exclusive probability of a click is always less or equal than the inclusive one, which for $k\neq l$ implies that
\begin{equation}
q  \le \bigl\langle 1 - e^{-\mu^{(l)}_{k}}  \bigr\rangle \le \langle \mu^{(l)}_{k} \rangle = \bar{n}(1-e^{-\sigma^2}).
\label{Eq:qapprox}
\end{equation}
The above bounds are conservative, underestimating the probability of a correct identification of the Hadamard sequence while overestimating the probability of an identification error.

\section{Optimization}
\label{Sec:Optimization}

\begin{figure}
   \centering
   \includegraphics[width=\columnwidth]{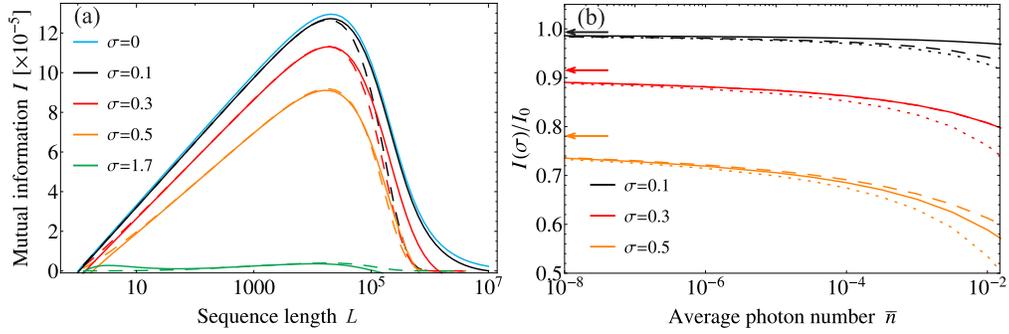}
    \caption{(a) Shannon mutual information $I$ given in Eq.~(\ref{Eq:Rcomplete}) as a function of the sequence length $L$ for $\bar{n}=10^{-5}$ and several dephasings $\sigma$ calculated for $p$ and $q$ estimated in Eqs.~(\ref{Eq:papprox}) and (\ref{Eq:qapprox}) (solid lines) and using Eq.~(\protect\ref{Eq:pjlgeneral}) expanded up to $\bar{n}^2$ (dashed lines). The case $\sigma = 1.7$ illustrates the breakdown of approximations used in the analysis. (b) The ratio $I(\sigma)/I_0$ characterizing the effect of phase fluctuations on mutual information as a function of the average photon number $\bar{n}$. Information $I(\sigma)$ is optimized over the sequence length $L$ treated as a continuous parameter taking bounds given in Eq.~(\ref{Eq:papprox}) and (\ref{Eq:qapprox}) (solid lines), numerical expansions up to $\bar{n}^2$ (dashed lines), and the closed approximate formula from Eq.~(\ref{Eq:Iclosedanalytical}) (dotted lines). The reference value $I_0$ has been obtained by optimizing numerically over $L$ the exact expression for mutual information in the noiseless case. Arrows indicate asymptotic values $e^{-\sigma^2}$. For $\sigma=0.3$ dashed and solid lines overlap.}
    \label{Fig:RofL}
\end{figure}

Starting from the standard definition \cite{CoverThomas}, a lengthy but straightforward calculation shows that Shannon mutual information per time bin for a generalized erasure channel with uniform noise can be written as
\begin{equation}
I = \frac{p}{L} \log_2 L
- \frac{1}{L} [p+(L-1)q] H \left( \frac{p}{p+(L-1)q} \right)
- q \left(1-\frac{1}{L}\right )\log_2 \left(1-\frac{1}{L}\right ) ,
\label{Eq:Rcomplete}
\end{equation}
where $H(x) = - x \log_2 x - (1-x) \log_2 (1-x) $ denotes binary entropy. In Fig.~\ref{Fig:RofL} we present $I$ as a function of the
sequence length $L$ for $\bar{n}=10^{-5}$ and increasing dephasing  calculated using two approaches. The first approach is to compute numerically the probabilities using Eq.~(\ref{Eq:pjlgeneral}) expanded up to the second order in $\bar{n}$, while the second one uses bounds derived in Eqs.~(\ref{Eq:papprox}) and (\ref{Eq:qapprox}). It is seen that both calculations yield very close locations of the optimal sequence length $L$. Furthermore, the optimal $L$ does not change noticeably with the strength of dephasing as long as the latter does not become too large. One should also note that the permitted values of $L$ defined by the dimensionality of Hadamard matrices are evenly distributed in the logarithmic scale used for the abscissa in Fig.~\ref{Fig:RofL}(a). Following this observation, we will treat $L$ as a continuous parameter when optimizing mutual information.

Let us now inspect closer the expression for mutual information given in Eq.~(\ref{Eq:Rcomplete}) with the aim to derive an approximate analytical formula for its optimal value in the regime of weak dephasing. Motivated by the numerical example shown in Fig.~\ref{Fig:RofL}(a), we will assume that $L \gg 1$. The self-consistency of this assumption will be confirmed afterwards.
It is easy to verify that the factor $[p+(L-1)q]/L$ multiplying the binary entropy in the second term is equal to $\bar{n}$ in the leading order, whereas the last term in Eq.~(\ref{Eq:Rcomplete}) is of the order $O(\bar{n}/L)$ and therefore can be neglected for long sequences. Further,
applying  a linear expansion in $\bar{n}$ to $p$ and $q$ in the second term of Eq.~(\ref{Eq:Rcomplete}), using Eq.~(\ref{Eq:avgmu}), and taking $L \gg 1$, mutual information can be simplified to an approximate form $I \approx (p/L) \log_2 L - \bar{n} H(e^{-\sigma^2})$. Explicit dependence on $L$ occurs only in the first term,  which has a form
analogous to mutual information for a standard erasure channel with the non-erasure probability equal to $p$.

Given this analogy, we will use the approach presented in \cite{Jarzyna} and expand $p$ up to the second order
in $L\bar{n}$:
\begin{equation}
p \approx e^{-\sigma^2} L \bar{n} - {\textstyle \frac{1}{2}} (2e^{\sigma^2} -1) (e^{-\sigma^2} L \bar{n})^2.
\end{equation}
This makes our problem equivalent to optimizing the PPM format when the average photon number per bin is equal to $e^{-\sigma^2}\bar{n}$ and the non-zero pulses exhibit super-Poissonian photon statistics characterized by the normalized second-order intensity correlation function $g^{(2)} = 2e^{\sigma^2} -1$. The solution to this problem found in \cite{Jarzyna} satisfies assumptions made about the optimal sequence length $L_{\text{opt}}$, i.e.\ $\bar{n} \ll L_{\text{opt}} \bar{n} \ll 1$. The approximate expression for mutual information based on this result can be compactly written as
\begin{equation}
I(\sigma) \approx e^{-\sigma^2} \bar{n} \Pi \bigl( (2-e^{-\sigma^2}) \bar{n} \bigr) - \bar{n} H(e^{-\sigma^2}),
\label{Eq:Iclosedanalytical}
\end{equation}
where $\Pi(x)$ is the photon information efficiency function for the PPM format with Poissonian pulse photon statistics given explicitly in \cite{Jarzyna}.

Figure~\ref{Fig:RofL}(b) depicts the ratio $I(\sigma)/I_0$  of mutual information optimized over the sequence length in the presence of phase noise $I(\sigma)$ to the ideal noiseless case $I_0$ as a function of the average photon number $\bar{n}$. Two approaches used to produce Fig.~\ref{Fig:RofL}(a) are found in good agreement with the closed analytical expression derived in Eq.~(\ref{Eq:Iclosedanalytical}), which confirms the consistency of our analysis. Using in Eq.~(\ref{Eq:Iclosedanalytical}) the asymptotic form of the photon information efficiency function
$\Pi(x) \approx \log_2 (1/x)$ for small arguments $x\ll 1$ yields in the leading order $I(\sigma) \approx e^{-\sigma^2} \bar{n}\log_2(1/\bar{n})$. Consequently, the ratio $I(\sigma)/I_0$ tends for $\bar{n} \rightarrow 0$ to a constant value $e^{-\sigma^2}$, marked with arrows in the ordinate of Fig.~\ref{Fig:RofL}(b). It is seen that this asymptotic value
is attained faster with decreasing $\bar{n}$ for weaker dephasings $\sigma$. Importantly, the Holevo-type scaling of mutual information in the presence of phase noise is retained and the principal effect of phase fluctuations in the asymptotic regime $\bar{n} \rightarrow 0$ is introduction of a multiplicative  factor $e^{-\sigma^2}$.

\section{Conclusions}
\label{Sec:Conclusions}

The communication scheme based on Hadamard words constructed from BPSK symbols with a linear-optics joint-detection receiver\cite{Guha} attains superadditive accessible information through collective measurements of transmitted optical signals. Quantum-enhanced detection schemes are potentially more sensitive to noise and imperfections, as recently noted in the case of individual quantum receivers for the BPSK format \cite{Olivares}. Theoretical analysis presented in this work shows that a moderate amount of phase noise can be tolerated in the collective BPSK scheme without compromising the Holevo-type scaling of mutual information with the mean photon number.

In our calculations, the phase variables fluctuated around a reference value that remained constant over the entire Hadamard sequence. An alternative model to analyze would be free-floating Markovian phase diffusion discussed recently in \cite{JarzynaJPA} for a different keying scheme. It should also be noted that there exist coherent strategies capable of reading out the PPM format more efficiently compared to direct detection \cite{ChenPPM}. More generally, it would be interesting to investigate practical schemes to attain collective enhancement in the case of larger alphabets such as quadrature phase shift keying for which individual quantum receivers operating below the shot noise limit have been recently demonstrated \cite{MullerQPSK,BecerraPSK}.

\section*{Acknowledgments}
We acknowledge insightful discussions with F. E. Becerra, R. Demkowicz-Dobrza\'{n}ski, and Ch. Marquardt.
This research was partly supported by the EU 7th Framework Programme projects SIQS (Grant Agreement No. 600645) and PhoQuS@UW (Grant Agreement No. 316244).
\end{document}